\documentclass[
]{ceurart}

\usepackage{subcaption}
\usepackage{listings}

\begin{document}

\copyrightyear{2026}
\copyrightclause{Copyright for this paper by its authors. Use permitted under Creative Commons License Attribution 4.0 International (CC BY 4.0).}

\conference{CSEDM’26: 10th Educational Data Mining in Computer Science Education (CSEDM) Workshop, July, 2026, Seoul, Korea}

\title{Automated Recommendation of Programming Learning Content Using Pattern-based Knowledge Components}

\author[1]{Muntasir Hoq}

\author[1]{Griffin Pitts}

\author[2]{Zhangqi Duan}

\author[3]{Arun Balajiee Lekshmi Narayanan}

\author[3]{Mohammad Hassany}

\author[2]{Andrew Lan}

\author[3]{Peter Brusilovsky}

\author[1]{Bita Akram}
\cormark[1]

\address[1]{North Carolina State University, Raleigh, NC, USA}
\address[2]{University of Massachusetts, Amherst, MA, USA}
\address[3]{University of Pittsburgh, Pittsburgh, PA, USA}
\cortext[1]{Corresponding author: Bita Akram, bakram@ncsu.edu}

\begin{abstract}
  Introductory programming instruction relies on hands-on practice and short learning activities to support mastery of foundational concepts. Although many such learning resources exist, organizing and linking these items in instructionally meaningful ways is challenging without time-intensive expert curation. This study investigates the use of pattern-based Knowledge Components (KCs) to automatically identify code-based learning resources targeting similar concepts. In our approach, pattern-based KCs are extracted from each code sample, and related activities are identified by measuring similarity between the KC sets associated with each activity. By leveraging alignment at the level of semantically important programming patterns, this method supports contextually appropriate and pedagogically useful recommendations. We evaluate our approach on an expert-organized corpus of introductory Python materials in which instructors grouped items into bundles based on conceptual similarity. Results show that our pattern-based KC approach retrieves resources that align with this expert organization, and outperformed representative KC- and embedding-based baselines across standard ranking evaluations. Overall, the framework supports targeted, concept-oriented guidance for programming learners and can help instructors organize, bundle, and recommend instructional content at scale.
\end{abstract}


\maketitle

\section{Introduction}
Example-based problem solving has shown promise for improving students' learning across educational domains~\cite{renkl2014toward}. In computer science (CS) education, one common activity type is a ``worked example''. In a worked example, a correct solution of a programming problem (i.e., code) is presented to students along with instructional explanations about each coding step that can be revealed as needed~\cite{hosseini2020improving,pitts2026personalized}. Timely exposure to a relevant example can help students deepen their understanding and progress when they encounter difficulty~\cite{hosseini2020improving,pitts2026personalized,najar2016}. Complementing worked examples, introductory programming environments often provide practice-oriented activities that challenge students to apply what they know, for instance, by completing incomplete code~\cite{hosseini2020improving}, or arranging code lines in a correct order~\cite{parson2006}. 

Prior work suggests that alternating between types of activities focusing on similar concepts can be beneficial for students, especially when semantic similarity between activities is curated by experts~\cite{akhuseyinoglu2024worked}. For example, a study using the Program Construction Examples (PCEX) system, bundling worked examples with closely related completion problems, increased engagement and improved students' performance~\cite{hosseini2020improving}. Subsequent studies confirmed that semantic similarity between a practice-oriented activity and its paired worked example is a driver of problem-solving success and persistence~\cite{akhuseyinoglu2024worked}.

Though an abundance of programming activities are available across repositories, it remains difficult to maintain links among activities that target similar programming concepts and provide learners with relevant items at the moment of need~\cite{barria2019need}. Instructors have traditionally addressed this by manually linking items, such as assigning practice problems with examples~\cite{Najar2014adaptive}. However, human curation does not scale well. In large repositories, maintaining these links becomes slow, difficult to update, and prone to inconsistency. Early automated approaches mostly used surface signals, such as keywords or n-grams~\cite{kibby1989towards,mayes1988strathtutor}. These features are computationally efficient, but they often match code that appears similar while targeting different underlying concepts, or fail to connect conceptually similar solutions expressed with different syntax. Knowledge-based approaches, such as ontology-based linking, improved relevance by indexing items with domain concepts and accounting for relationships among those concepts~\cite{hosseini2017study}. However, these methods typically require hand-engineered domain concept models and extensive expert annotation~\cite{carr2001conceptual,koedinger2012knowledge}. This motivates learning activity representations that are grounded in program structure and do not require extensive expert annotation.

Motivated by this gap, this paper presents a knowledge component (KC)-based recommendation framework that derives similarity from program representations in terms of the KCs they represent. We hypothesize that KCs, defined as the skills or concepts a learner must apply to solve a task, provide an appropriate level of abstraction for recommending instructional content across worked examples and practice activities~\cite{koedinger2012knowledge}. Matching at the KC level prioritizes items that reflect similar conceptual understanding required by a target activity. We adopt a pattern-based KC formulation~\cite{hoq2025pattern} in which KCs are recurring, semantically meaningful subtrees drawn from program Abstract Syntax Trees (ASTs). 

In our implementation, we discover candidate pattern-based KCs from a repository of programming activities using an explainable extraction method proposed in prior work~\cite{hoq2025pattern}. Each program is represented as an IDF-weighted knowledge vector over the resulting KC set, and candidate items are ranked using cosine similarity. The system can also project the highest-contributing KCs back into source code, yielding line-level rationales that allow instructors to audit why a recommendation was made. To evaluate recommendation quality, we use instructional content from PCEX~\cite{hosseini2020improving}, an online tool for interactive program learning. In PCEX, instructors organized worked examples and practice activities into topics and finer-grained bundles, which we treat as ground truth for assessing retrieval quality. We report Top-5 accuracy, mean reciprocal rank (MRR), and mean average precision (mAP) across two retrieval directions: retrieving worked examples for a given problem and retrieving problems for a given worked example. We also perform neighborhood tightness and clustering analyses to examine whether items from the same expert-defined bundle are placed closer together in the learned representation space. Across analyses, the pattern-based KC representation retrieves instructionally aligned items and compares favorably with code-similarity methods and alternative KC discovery baselines, including code2vec~\cite{alon2019code2vec}, SANN~\cite{hoq2025worked}, ontology-based approaches~\cite{hosseini2017study}, and recent LLM-identified KC approaches~\cite{duan2025automated,niousha2025llm,fan2025adaptive}. We further demonstrate the explainability of the proposed framework through a proof-of-concept example.

Our contributions are threefold: (i) a recommendation approach based on knowledge vectors that uses KC similarity across programming activities, (ii) an explainable pattern-based KC model that discovers and identifies programming patterns for recommendation, and (iii) an evaluation of how different KC models improve recommendation accuracy and alignment with expert-defined bundles.

\section{Related Work}
Early efforts to connect related learning materials emerged in educational hypertext under similarity-based navigation, where items were often matched by keyword overlap \cite{kibby1989towards,mayes1988strathtutor}. Later work introduced richer semantic approaches, often termed intelligent linking, including methods that compute similarity over metadata~\cite{tudhope1997navigation}. Related work on ontology-based linking indexed resources with ontology terms and computed similarity from their positions and relations within an ontology graph~\cite{carr2001conceptual,crampes2000ontology,dolog2003logic}.

In CS education, similarity estimation has often focused on program content. Gross et al.~\cite{gross2014select}, for example, linked Java materials using similarity between full-program abstract syntax trees (ASTs), though whole-program AST comparison can miss fine-grained correspondences in smaller structural units~\cite{weber1994elm}. Concept-based approaches address this limitation by representing problems and examples as vectors over domain concepts and computing similarity between these profiles~\cite{hosseini2014example}. Subsequent work extended this direction through ontology-informed similarity metrics for programming items~\cite{hosseini2017study} and recommendation methods for supplementing introductory programming textbooks with reusable learning content~\cite{barria2022augmenting}. However, automatically allocating and sequencing such resources remains challenging, especially when course ordering varies across instructors~\cite{sabet2022enriching}.

Recent advances in learned code representations have opened a new avenue for automatically extracting syntactic and semantic structure from programs using deep learning models \cite{hoq2023sann}. These representations reduce reliance on manual expert design of isomorphic problem-example pairs by enabling similarity search directly in the learned embedding space. Prior work shows that such methods can support a range of code-analysis tasks, including recommending similar learning content~\cite{hoq2024detecting,hoq2025worked}. However, these approaches primarily capture surface- or structure-level similarity between programs. In this work, we propose a KC-based recommendation framework grounded in the programming patterns required to solve a task. We evaluate this pattern-based KC framework against both code-similarity approaches and alternative KC representations.


\section{Dataset}

This work uses a Python programming dataset collected from the PCEX system~\cite{hosseini2020improving}. PCEX provides online access to worked code examples and short code completion tasks (``challenges''). Domain experts have organized these items into ``\textbf{bundles}'': compact groups of challenges and examples that target the same programming constructs and recurring patterns. Prior studies have validated expert-defined bundles~\cite{akhuseyinoglu2024worked,hosseini2020improving}, noting the value of aligning practice and assessment around shared, recurring programming patterns. Although constructing such bundles is time- and expertise-intensive, this limitation motivates the need for approaches that can support or scale this alignment.

\begin{figure}[t]
\centering

\begin{subfigure}[t]{0.39\linewidth}
    \centering
    \includegraphics[width=.9\linewidth]{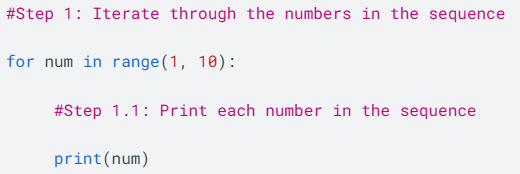}
    \caption{Same bundle code (Ex. 1)}
    \label{fig:bundle_ex1}
\end{subfigure}
\hfill
\begin{subfigure}[t]{0.39\linewidth}
    \centering
    \includegraphics[width=.9\linewidth]{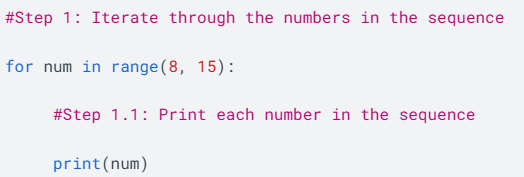}
    \caption{Same bundle code (Ex. 2)}
    \label{fig:bundle_ex2}
\end{subfigure}

\medskip

\begin{subfigure}[t]{0.35\linewidth}
    \centering
    \includegraphics[width=.9\linewidth]{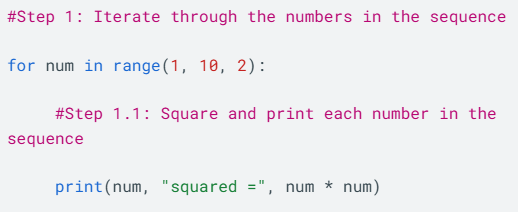}
    \caption{Different bundle code (Ex. 3)}
    \label{fig:bundle_ex3}
\end{subfigure}

\caption{Examples of recommendations with shared and different bundle-code patterns.}
\label{fig:bundle_examples}
\end{figure}

In PCEX, the dataset contains $123$ programs (examples and challenges) spanning $13$ topics, including Variables and Operations, If–Else, Boolean Expressions, For Loops, and Nested Loops. These items are partitioned into $49$ bundles (approximately four bundles per topic). A typical bundle consists of one worked example and one to three closely related challenges for the students; on average, each bundle includes $1.35$ challenges. We treat the existing topic and bundle assignments in PCEX, produced by expert instructors, as gold-standard labels for evaluating recommendation and clustering quality~\cite{sabet2022enriching}. Figures~\ref{fig:bundle_ex1} and~\ref{fig:bundle_ex2} illustrate two examples from the same bundle within the ``For Loops'' topic, while Figure~\ref{fig:bundle_ex3} shows an example from a different bundle in the same topic, emphasizing the structural distinctions that motivate bundle-level grouping.

\section{Methodology}
\begin{figure*}[t]
  \centering
  \includegraphics[width=0.66\textwidth]{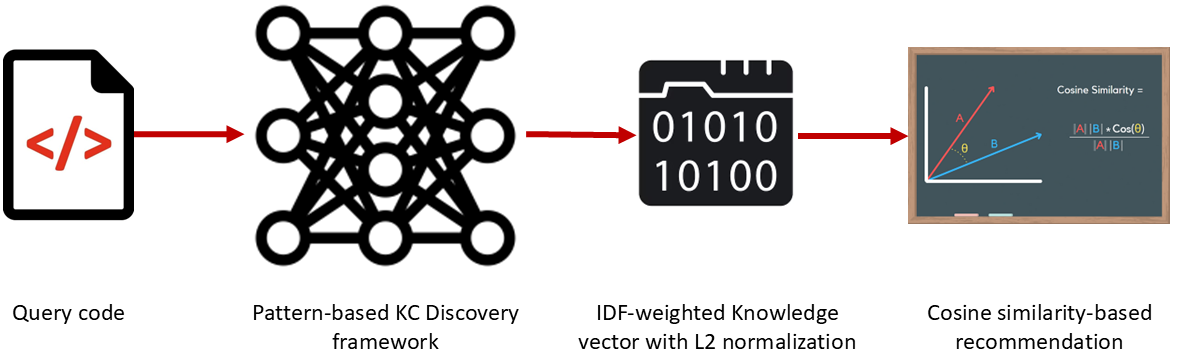}
  \caption{Learning activity recommendation pipeline.}
  \label{fig:method}
\end{figure*}

In this section, we describe our pipeline for recommending programming learning activities by matching items based on the patterns expressed in their associated code. We first extract \emph{pattern-based} KCs from each activity’s code sample. Each activity is then mapped to an IDF-weighted knowledge vector, and recommendations are produced by cosine similarity, supporting retrieval, as illustrated in Figure~\ref{fig:method}. We apply the KC discovery framework of~\cite{hoq2025pattern} to code artifacts in the PCEX dataset, producing (i) a set of structural KCs and (ii) a mapping from each learning activity to the KCs it expresses. 

\subsection{Pattern-based Knowledge Component Discovery}

Automatically discovering KCs from code is not a trivial task. In programming, small syntactic edits can obscure semantic equivalence, the same conceptual skill can be realized through many structurally distinct implementations, and the instructional relevance of a code fragment often depends on its surrounding program context rather than the fragment in isolation~\cite{shi2023kc}. 

To address this, we adopt pattern-based KCs~\cite{hoq2025pattern}, which define KCs as recurring, semantically meaningful structural patterns based on important subtrees from program Abstract Syntax Trees (ASTs). These subtree patterns capture programming constructs that recur across solutions, and their correct implementation is crucial for completing the assignments~\cite{hoq2025pattern}. Compared with KC models based on named concepts or curricular tags, pattern-based KCs provide finer-grained, implementation-level signals (e.g., guarded loop traversals, accumulator updates, boundary checks). This granularity is well matched to recommendations, since it allows the system to retrieve items that share structural patterns while also supporting line-level explanations of why an item was retrieved. Four steps of pattern-based KC discovery are specified below.

First, \textit{subtree pattern extraction}. Each program is converted into an AST using the Python AST parser\footnote{\url{https://docs.python.org/3/library/ast.html}}. Subtrees are extracted at multiple granularity levels, and a modified Subtree-based Attention Neural Network (SANN) \cite{hoq2025automated} assigns an attention weight to each subtree. These weights identify structural fragments that most strongly represent topic-relevant logic, and the most important subtrees per program are retained as candidate patterns. We train SANN using only coarse-grained topic labels for challenges and examples, deliberately withholding expert-defined bundle information. Although bundles encode finer-grained structure, our aim was to test whether the KC discovery framework can recover bundle-level regularities from coarser supervision through learned pattern similarity. This ensures that the extracted patterns do not trivially encode bundle membership, but reflect recurring structural regularities that generalize across problems within a topic. Data is split into $80$\% training, $10$\% validation, and $10$\% test sets, ensuring that every expert-defined bundle appears in the training split to preserve structural variation. Subtree embeddings use $64$ dimensions (selected from {$64$, $128$, $256$}). Models are trained with batch size $32$, up to $200$ epochs, and early stopping with patience $20$. Remaining hyperparameters follow prior work \cite{hoq2025pattern}.

Second, \textit{normalization}. We abstract selected subtrees by anonymizing identifiers, literal values, and other incidental tokens. We also canonicalize simple syntactic alternatives, such as commutative operator order, so that surface-different but conceptually equivalent code maps to the same abstract pattern. This step helps ensure that pattern-based KCs reflect conceptual similarity instead of lexical similarity. Introductory programming problems often admit multiple correct implementations that differ in variable naming, literal values, or syntactic choices, none of which should affect KC attribution.

Third, \textit{context-aware encoding}. While individual subtree patterns capture local structural information, their instructional meaning often depends on how they are composed within a complete program. For example, the same conditional structure may represent a boundary check, a classification decision, or part of an accumulator update, depending on its surrounding logic. To model these dependencies, we encode sequences of normalized subtrees using a context-aware latent representation. A $\beta$-variational autoencoder ($\beta$-VAE), using an LSTM, is used to encode the normalized subtrees to preserve how patterns co-occur within complete programs and across submissions, and learn a latent space in which structurally and semantically similar patterns lie close together. The $\beta$-VAE is trained to encode representative patterns for the important subtrees, guided by attention weights extracted from SANN. Following~\cite{hoq2025pattern}, the $\beta$-VAE model employed subtree embeddings of dimension $64$ and a latent space of dimension $128$, with intermediate dense layers matching the embedding size. To mitigate overfitting, a dropout rate of $0.2$ was applied. Training was performed for up to $50$ epochs using the Adamax optimizer with a learning rate of $0.001$, and early stopping with a patience of $10$ epochs. We adopted a $\beta$-VAE objective with a KL-divergence weighting factor of $\beta = 1 \times 10^{-2}$ to balance reconstruction accuracy and latent space regularization.
 
Fourth, \textit{clustering into KCs}. We cluster the latent vectors of the retained subtrees (with an attention threshold of 20\%, following~\cite{hoq2025pattern}) from programs using K-means (with the number of clusters selected by the silhouette heuristic). The resulting clusters form our pattern-based KCs, with centroids serving as canonical representatives and links preserved back to the original AST spans for explainability. Next, we use these KC assignments to compute similarity between learning activities for recommendation.

\subsection{Knowledge Vectors and Recommendation}

After discovering pattern-based KCs, we create a \emph{knowledge vector} that captures the presence of each KC for each code sample. For a corpus with \(K\) KCs, each program \(d\) is mapped to a binary incidence vector \(\mathbf{x}_d \in \{0,1\}^K\) indicating KC presence. We then apply an inverse-document-frequency (IDF) reweighting to emphasize discriminative patterns and downweight ubiquitous ones. Let \(N\) be the number of programs and \(n_i\) the number of programs that contain KC \(i\). The weight for KC \(i\) is
\[
w_i \;=\; \log\!\left(\frac{N}{n_i}\right).
\]
The resulting weighted representation is \(\mathbf{v}_d = \mathbf{x}_d \odot \mathbf{w}\), which we L2-normalize, so that dot products correspond to cosine similarity:
\[
\hat{\mathbf{v}}_d \;=\; \frac{\mathbf{v}_d}{\left\lVert \mathbf{v}_d \right\rVert_{2}}.
\]
Given a target item \(q\) (challenge or worked example), we rank candidate items \(e\) by cosine similarity
\[
s(q,e) \;=\; \hat{\mathbf{v}}_q^{\top}\,\hat{\mathbf{v}}_e,
\]
and return the top-ranked items as recommendations. Because similarity is computed in the pattern-based KC latent space, retrieval favors candidates that instantiate similar underlying program patterns expressed in \(q\), rather than items that are merely lexically or syntactically similar. 

We evaluate this approach on two retrieval tasks using the held-out test set: (1) retrieving worked examples for a target challenge and (2) retrieving challenges for a target worked example. A retrieved item is considered relevant if it belongs to the \emph{same expert-defined bundle} as the target item (our gold standard), with self-matches prevented. We report Top-5 accuracy, mean reciprocal rank (MRR), and mean average precision (mAP) to assess whether pattern-based KC-level similarity recovers bundle-aligned items relative to baselines.


\subsection{Baseline Approaches}
As baselines, we compare against (1) code2vec-based code vectorization~\cite{alon2019code2vec}, (2) ontology concept-based code  vectorization~\cite{hosseini2017study}, (3) SANN-based code vectorization~\cite{hoq2025worked}, (4) KCGEN-KT KCs~\cite{duan2025automated}, and (5) LLM-KCI KCs \cite{niousha2025llm,fan2025adaptive}. 

\textbf{code2vec.} code2vec learns a vector representation of a code snippet by decomposing its abstract syntax tree into many “path-contexts.” Each path-context is formed by two terminal tokens together with the AST path connecting them, written as (token\_s, path, token\_t). The model embeds the tokens and the path and then uses an attention mechanism to weight and aggregate the set of path-context embeddings into a single fixed-length vector for the snippet.

\textbf{Ontology-KC.} Ontology-based concept vectorization represents programming items using concept annotations derived from their source code. Following prior work~\cite{hosseini2017study}, a set of high-level programming concepts from each code submission was extracted using an open source static code parser\footnote{\url{https://acos.cs.aalto.fi/python-parser}} using an expert-defined programming ontology\footnote{\url{https://acos.cs.aalto.fi/static/python-parser/ontology.png}}. Each submission is then encoded as a concept-based vector, and similarity between submissions is computed based on cosine similarity.

\textbf{SANN.} We treat SANN as a standalone code-embedding model and use only its learned source-code vectors derived from AST structure (no topic, bundle, or knowledge signals beyond what is implicit in the encoder). Each challenge and worked example is embedded once with SANN, and recommendations are produced by cosine similarity in this SANN embedding space, following~\cite{hoq2025worked}.

\textbf{KCGen-KT.} KCGen-KT~\cite{duan2025automated} represents each item with a set of LLM-generated KCs. It prompts GPT-4o to propose candidate KCs from the problem statement and canonical solution, clusters semantically similar candidates into KC groups, and summarizes each cluster into a consolidated KC description, yielding 30 unique KCs. These KCs are assigned as tags to items to form a KC-based representation.

\textbf{LLM-KCI.} LLM-KCI~\cite{niousha2025llm} indexes programming items by inducing an expert-curated KC inventory with an LLM and then using that inventory to assign KCs to items. It prompts GPT-4o to generate candidate KCs from the available item context, aggregates and removes semantic duplicates to form a KC list, and then prompts the LLM with this list to select the most relevant KCs for each item, yielding a sparse KC assignment vector used for similarity-based retrieval.

\section{Experiments and Results}
To evaluate whether our pattern-based KC representation supports accurate recommendations of programming learning activities, we first examine the alignment of the knowledge vector using pattern-based KCs with expert-defined topics and bundles using tightness analyses. As described in Section 3, we evaluate our recommendation approach on the PCEX instructional repository~\cite{hosseini2020improving}. We test two retrieval directions that reflect common instructional needs, recommending worked examples for a target challenge activity and recommending challenge activities for a target worked example. We treat expert-defined bundle membership as ground truth and report Top-5 accuracy, mean reciprocal rank (MRR), and mean average precision (mAP). We also illustrate explainability with visualizations that project the highest-contributing pattern-based KCs back onto code to explain specific recommendations, and include an ablation on IDF weighting to isolate its contribution.

\subsection{Knowledge Vector Alignment with Expert-Defined Topics and Bundles}


We use expert-defined bundles as our ground truth and hypothesize that these bundles capture fine-grained similarity among challenges and worked examples. Representations derived from the knowledge vectors should therefore place items from the same bundle closer together than items from different bundles. The same intuition extends to topics at a coarser granularity: items within a topic should be nearer to one another than to items from other topics, but not as tightly grouped as items within a bundle. To test this hypothesis, we quantify tightness as the mean pairwise cosine distance among vectors within a group, with lower values indicating tighter, more coherent groups. Using expert labels as our standard, we computed tightness separately for each bundle and each topic. 

Bundles consistently exhibit smaller average distances than topics, indicating more compact clusters at the bundle level. The average bundle tightness was $0.51$ (cosine-distance units), compared to $0.81$ for topics, showing that bundle members are considerably closer to one another than items that merely share a topic. The dataset-wide average distance was $1.7$, reflecting the broader diversity across programs and reinforcing that bundle membership captures highly similar code patterns, whereas topic membership reflects a looser but still meaningful similarity.



\begin{table*}[t]
\centering
\caption{Performance of different KC modeling approaches for recommending worked examples and challenges. Higher is better for all metrics.}
\label{tab:kc-recommendation}
\setlength{\tabcolsep}{7pt}
\renewcommand{\arraystretch}{1.15}
\begin{tabular}{lccc ccc}
\toprule
& \multicolumn{3}{c}{Worked-example retrieval} & \multicolumn{3}{c}{Challenge retrieval} \\
\cmidrule(lr){2-4}\cmidrule(lr){5-7}
Model & Top-5 & MRR & mAP & Top-5 & MRR & mAP \\
\midrule
LLM-KCI            & 0.87 & 0.79 & 0.79 & 0.88 & 0.78 & 0.78 \\
KCGen-KT KC        & 0.84 & 0.62 & 0.62 & 0.85 & 0.64 & 0.65 \\
Ontology-based KC  & 0.88 & 0.67 & 0.65 & 0.89 & 0.66 & 0.65 \\
code2vec embeddings& 0.83 & 0.71 & 0.71 & 0.82 & 0.70 & 0.71 \\
SANN embeddings    & 0.87 & 0.78 & 0.78 & 0.87 & 0.77 & 0.78 \\
\textbf{Pattern-based KC}
                  & \textbf{0.89} & \textbf{0.81} & \textbf{0.82}
                  & \textbf{0.90} & \textbf{0.80} & \textbf{0.79} \\
\bottomrule
\end{tabular}
\end{table*}

\subsection{Clustering Bundles without Expert Effort}

To assess whether the knowledge vectors recover the dataset’s latent bundle organization without manual expert curation, we performed agglomerative hierarchical clustering with cosine distance and complete linkage on the knowledge vectors. The dendrogram preserved the original pairwise geometry well (cophenetic correlation, $r$=$0.819$. Values closer to 1 mean the tree faithfully reflects the distances). We then evaluated alignment with expert bundlings \emph{post hoc} by sweeping distance cut thresholds and computing Adjusted Rand Index (ARI) and V-measure (the harmonic mean of homogeneity and completeness) for the induced flat partitions. At the \emph{bundle} level, the best ARI was $0.605$ at $0.6$ threshold with $63$ clusters, indicating substantial agreement beyond chance. The accompanying V-measure was $0.933$ (homogeneity $0.949$ and completeness $0.916$), suggesting clusters were both internally pure and capture most items of each expert bundle.


Additionally, we quantified local neighborhood structure by identifying, for every program, its single nearest neighbor in the clustering (excluding self-matches) and categorizing the match as: (1) \emph{same bundle}, (2) \emph{same topic but different bundle}, or (3) \emph{different topic}. The distribution of nearest-neighbor types was: $76.4$\% same bundle, $11.9$\% same topic/different bundle, and $11.7$\% different topic. The high same-bundle rate indicates that knowledge vectors place structurally similar problems and examples in close proximity, while same-topic matches capture looser but meaningful commonalities across bundles within a topic. Further inspection of mismatches indicates that they often reflect broadly used programming patterns, such as common control-flow in Loops, Nested Loops, and Strings, that legitimately span topic boundaries. This analysis shows that knowledge vectors can automatically recover bundle-level groupings within topics and expose finer substructure, reducing the need for manual pre-linking of learning content.

\subsection{KC-based Recommendation of Learning Content}

We first report retrieval performance under the KC-based similarity model. For each target item, we rank candidate items by cosine similarity of their knowledge vectors. A recommendation is counted correct when at least one of the top five retrieved items belongs to the same expert-defined bundle as the target item. Table \ref{tab:kc-recommendation} reports Top-5 accuracy, mean reciprocal rank (MRR), and mean average precision (mAP) for both retrieval directions across all baselines.

The pattern-based KC approach achieves the strongest retrieval performance in both directions, indicating that our KC-based retrieval method, which combines pattern-based KCs with IDF-weighted knowledge vectors, more reliably ranks bundle-aligned items ahead of learned code embeddings and other KC-based baselines. Improvements in MRR and mAP further suggest that relevant items tend to appear earlier in the ranking, which is important in practice because instructional systems typically surface only the top one or two recommendations to learners.

For worked-example retrieval, the pattern-based KC approach achieves Top-5 accuracy of $0.89$\%, MRR$=0.81$\%, and mAP$=0.82$\%, outperforming code2vec embeddings (Top-5$=0.83$\%, MRR$=0.71$\%, mAP$=0.71$\%), ontology-based KCs (Top-5$=0.88$\%, MRR$=0.67$\%, mAP$=0.65$\%), and the remaining baselines. For challenge retrieval, the pattern-based KC approach again ranks best with Top-5$=0.90$\%, MRR$=0.80$\%, and mAP$=0.79$\% (Table~\ref{tab:kc-recommendation}).


We also conducted an ablation that removes IDF reweighting when constructing knowledge vectors for our pattern-based KC approach. In this setting, each program is represented as a binary KC-incidence vector indicating which KCs are present. Without IDF, worked-example retrieval achieves Top-5 accuracy of $88$\%, mAP of $72$\%, and MRR of $76$\%, and the challenge retrieval achieves Top-5 accuracy of $85$\%, mAP of $68$\%, and MRR of $75$\%. Compared with the IDF-weighted knowledge vectors, IDF yields a +$1$\% percentage point (pp) gain in Top-5 accuracy for worked-example retrieval ($89$\% vs. $88$\%), a +$5$\% pp gain in MRR ($81$\% vs. $76$\%), and a +$10$\% pp gain in mAP ($82$\% vs. $72$\%). For challenge retrieval, IDF yields a +$5$ pp gain in Top-5 accuracy ($90$\% vs. $85$\%), a +$5$\% pp gain in MRR ($80$\% vs. $75$\%), and a +$11$ pp gain in mAP ($79$\% vs. $68$\%), supporting the value of down-weighting ubiquitous KCs. 



\subsection{Explainability of KC–based Recommendations}

A benefit of recommending at the level of programming patterns is that the rationale for a match is explicit, with pattern-based KCs emphasizing instructionally important patterns. For example, any targeted \emph{challenge} \(q\) and candidate \emph{example} \(e\), their knowledge vectors \(\mathbf{v}_q\) and \(\mathbf{v}_e\) decompose the cosine score into per–KC contribution, $s(q,e)$, as:
\[
s(q,e)=\hat{\mathbf{v}}_q^{\top}\hat{\mathbf{v}}_e=\sum_{i=1}^{K}\;\hat{v}_{q,i}\,\hat{v}_{e,i},
\]

The KCs with the largest products \(\hat{v}_{q,i}\hat{v}_{e,i}\) are the primary drivers of the recommendation. To visualize this rationale, we trace each high-contributing KC back through the extraction pipeline to the high-attention subtree instances associated with that KC, then map those instances to their code spans. We highlight these spans with color intensity proportional to the underlying attention weight. This provides a line-level view of \emph{which patterns} influenced the match, without implying that raw AST subtrees are directly clustered or compared during retrieval. Prior work has validated that these high-attention subtrees capture instructionally meaningful evidence related to program correctness and logical errors, supporting their use as explanatory units \cite{hoq2025automated}.

\begin{figure}[t]
  \centering

  \begin{subfigure}[t]{0.45\linewidth}
    \centering
    \fbox{\includegraphics[width=.82\linewidth]{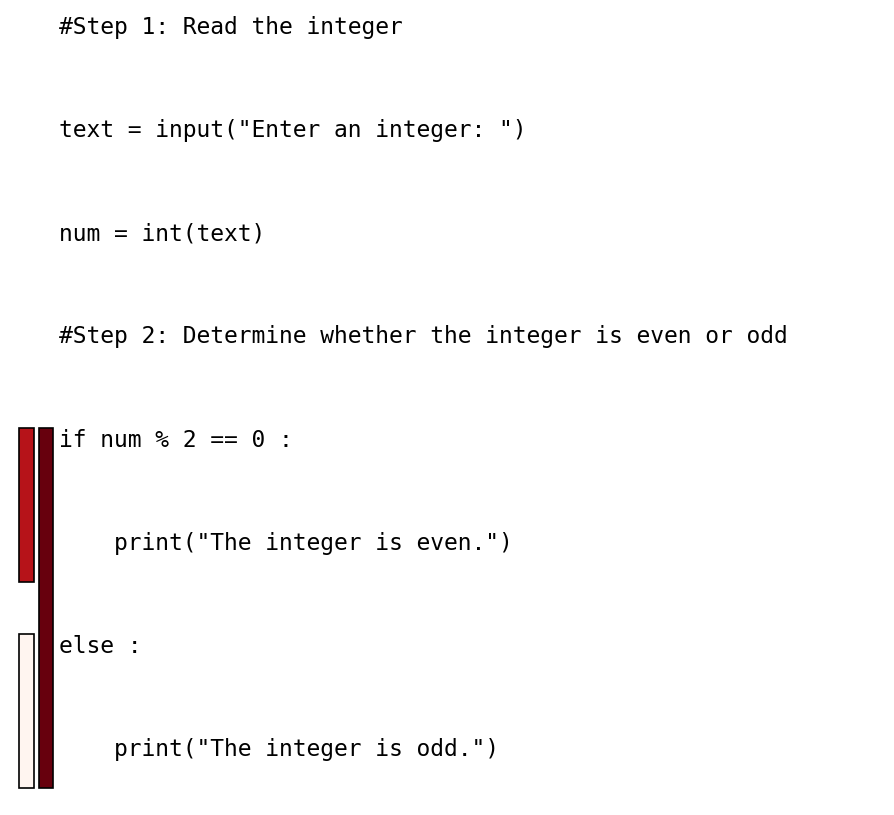}}
    \caption{An example challenge code.}
    \label{fig:challenge_vis}
  \end{subfigure}
  \hfill
  \begin{subfigure}[t]{0.45\linewidth}
    \centering
    \fbox{\includegraphics[width=.9\linewidth]{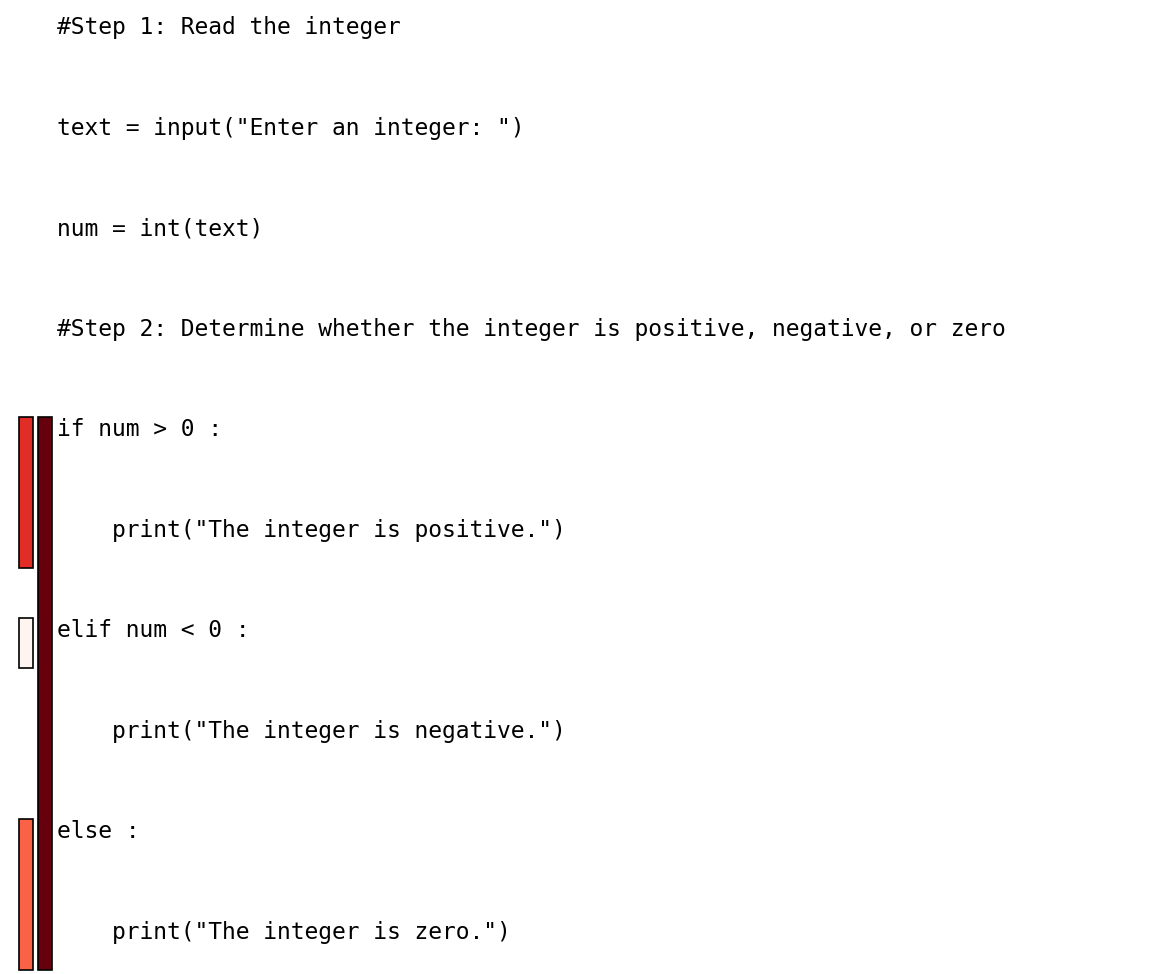}}
    \caption{Recommended worked example.}
    \label{fig:example_vis}
  \end{subfigure}

  \caption{KC-level similarity within the \emph{If-Else} topic from the same bundle. Left: challenge code that determines whether an input integer is \emph{even or odd}; right: recommended worked example that determines whether an input integer is \emph{positive, negative, or zero}. Colored rails in the left margin mark the top-contributing pattern-based KCs; color intensity is proportional to the attention weight.}
  \label{fig:challenge_example}
\end{figure}

Figures~\ref{fig:challenge_vis} and~\ref{fig:example_vis} illustrate an explanation from the same bundle within the \emph{If-Else} topic. In the challenge (left), attention concentrates on the binary conditional structure used to determine whether an input integer is even or odd, highlighting the modulus-based predicate and the associated if–else branching logic. In the recommended worked example (right), attention emphasizes the conditional decision structure used to classify an integer as positive, negative, or zero, which involves a multi-branch if–elif–else construct. Although the specific predicates and the number of branches differ, both instantiate the same underlying control-flow schema—sequential input processing followed by mutually exclusive conditional evaluation and branching—explaining why the worked example is retrieved. In contrast, scaffolding elements such as input prompts and output statements receive little or no highlight.

These explanations are designed to support \emph{pattern-focused learning}. For students, the highlighted regions direct attention to the abstract decision-making schema underlying conditional logic, such as evaluating a predicate and selecting among mutually exclusive branches. This encourages learners to internalize the general structure of conditional reasoning and transfer it across problems. For instructors, KC-level contribution breakdowns provide an auditable account of why a particular example was retrieved, help diagnose failure modes, and support inspection and maintenance of the KC inventory and content repository.

\section{Discussion}

This work addresses a challenge in programming education: recommending learning resources at scale that are contextually relevant to a student’s current learning focus \cite{brusilovsky2003adaptive}. Our approach tags learning resources with pattern-based KCs that capture instructionally important programming patterns in students’ code, enabling more targeted recommendations. By aligning retrieval with pedagogical units such as KCs, the recommended resources are likely to address the skills students need to practice. 

To assess whether the pattern-based KC vectors capture educationally meaningful structure, we conducted a tightness and clustering analysis. The tightness results show clear alignment with the expert-bundled content: learning activities within the same instructor-defined bundle form substantially tighter neighborhoods than those grouped only at the topic level, while topic-level groupings remain more compact than the dataset as a whole. This ordering is consistent with instructional intent, since bundles are designed to capture near-isomorphic problem families, suggesting that the representation encodes fine-grained, instructionally relevant similarity.

We then used hierarchical clustering as an unsupervised validation of the representation space. This analysis examined whether learning activities naturally organize into coherent groups based only on their knowledge vectors, without access to expert-defined bundle labels. In contrast to retrieval-based evaluation, which depends on ranked neighbor lists and top-N cutoffs, clustering tests whether programs that require similar KCs are placed near one another. The strong alignment between induced clusters and expert bundles indicates that pattern-based KC vectors capture conceptual similarity within the programs. This recovery of bundle-level structure suggests that the approach may generalize to new or unlabeled repositories. Such clustering could support instructors by proposing candidate bundles, auditing existing groupings, or surfacing related activities that share underlying patterns across topics.

Experimental results further showed that the proposed framework achieves strong retrieval performance on the PCEX dataset when evaluated against expert-annotated bundles as ground truth. Using KC-level similarity, the method performs well under rank-based metrics, including Top-$5$, MRR, and mAP, and it outperforms concept vectors~\cite{hosseini2017study}, code2vec~\cite{alon2019code2vec}, and SANN~\cite{hoq2025worked} embedding, and LLM-based KC vectors~\cite{duan2025automated,niousha2025llm}. Compared with other embedding approaches, pattern-based KCs better capture semantic similarity by reducing reliance on surface-level information through normalization and improving generalization through VAE training. LLM-based KC identification approaches can map similar implementations of programming concepts, but may suffer from sparsity and limited explainability. LLM-KCI showed competitive performance when supplied with an instructor-defined KC set, but constructing such inventories requires substantial expert effort and may reflect an instructor’s subjective view of the domain. Pattern-based KCs are learned directly from student code, allowing the representation to adapt its granularity based on where students struggle and which patterns require mastery. Ontology-based KC approaches exhibit lower MRR and mAP, suggesting that predefined conceptual structures are less effective at capturing fine-grained patterns in student solutions.

We also demonstrated the explainability of our model, an important characteristic to increase the trust of instructors and educators alike \cite{pitts2025understanding}. As each pattern-based KC is anchored to AST spans, we can project top-contributing KCs back onto programs for both the targeted item and the recommendation, producing line-level rationales. These explanations help students focus on the specific constructs they need to review. They also let instructors audit the basis for a match, diagnose false positives, and refine the resource inventory over time. More broadly, this supports actionable analytics and interfaces, such as open learner model views~\cite{barria2017concept} that communicate progress and gaps at the KC level. 

Overall, the proposed pipeline offers a scalable alternative to manually curated links, which become costly as repositories grow \cite{hosseini2020improving}. By discovering and reusing structural KCs across items and topics, the system can generalize to new materials with limited expert effort, supporting maintenance and expansion of content libraries. This helps bridge the gap between large repositories and personalized support, complementing prior work that showed the value of bundling learning content \cite{hosseini2020improving}.


\section{Limitations and Future Work}

Although the proposed KC-based recommendation method shows strong performance, there are multiple avenues for improving the work in the future. First, the dataset is small ($123$ programs), which may constrain both pattern-based KC discovery and evaluation sensitivity. Future work should expand the dataset across additional course offerings, textbooks, and activity types, then measure how recommendation quality changes as the number and diversity of learning activities increase. Second, explanation quality depends on the attention weights assigned by the framework and the KC clustering resolution. We highlight KC patterns that are the primary drivers for a recommendation; however, highly overlapping KCs can produce adjacent or redundant highlights. Future work will (i) calibrate attention thresholds and clustering granularity, (ii) merge near-duplicate KCs, and (iii) validate explanation fidelity through instructor judgments. Third, the first step of the KC extraction pipeline uses topic labels for training, limiting direct application to unlabeled corpora. Although many instructional resources are organized by topic, future work should explore weakly supervised or self-supervised alternatives and examine performance when labels are incomplete or noisy. Finally, future classroom studies should measure impacts on problem-solving performance, time-to-solution, and transfer when recommendations and explanations are available. 

\section{Conclusion}

This paper presented a recommendation framework for suggesting relevant programming learning resources by matching code at the level of pattern-based knowledge components. Similarity analysis showed that the resulting knowledge vectors align with expert-defined topics and bundles, indicating that they capture fine-grained programming patterns reflected in instructional content organization. Retrieval results showed that the pattern-based KC approach produces bundle-aligned recommendations and outperforms representative KC- and embedding-based baselines. The framework also provides line-level rationales by highlighting the code regions responsible for a match, supporting transparency for students and instructors. Overall, the method provides a scalable alternative to manual curation of relevant programming activities by aligning recommendations with the underlying skills a program requires and, thus, supporting targeted remediation.

\begin{acknowledgments}
This research was supported by the National Science Foundation (NSF) under Grants \#2418658, \#2418655, and \#2418657. Any opinions, findings, and conclusions expressed in this material are those of the authors and do not necessarily reflect the views of NSF.
\end{acknowledgments}

\section*{Declaration on Generative AI}
  During the preparation of this work, the authors used ChatGPT in order to: Paraphrase and reword. After using this tool/service, the authors reviewed and edited the content as needed and take full responsibility for the publication’s content.
  
\bibliography{aaai2026}

\appendix

\end{document}